\documentstyle[aps,psfig,epsf]{revtex}

\begin{document}
\title{Alternative mechanism of the sign-reversal effect in
superconductor-ferromagnet-superconductor Josephson junctions.\\
}
\author{A.F. Volkov$^{1,2}$ and A. Anishchanka$^{1}$}
\address{$^{(1)}$Theoretische Physik III,\\
Ruhr-Universit\"{a}t Bochum, D-44780 Bochum, Germany\\
$^{(2)}$Institute of Radioengineering and Electronics of the Russian Academy%
\\
of Sciences, 103907 Moscow, Russia }
\maketitle

\begin{abstract}
We consider a simple model of a multidomain
superconductor-ferromagnet-superconductor (SFS) Josephson junction.
Sign-alternating magnetization $M$ in domains leads to a spatial modulation
of the phase difference $\phi (x)$. Due to this modulation the Josephson
critical current $I_{c}$ may have a different sign depending on the ratio of
the magnetic flux in a domain, 4$\pi Ma(2d_{F}),$ to the magnetic flux
quantum. This phase modulation, but not a nonmonotonic dependence of the
local critical current density $j_{c}$, may be the reason for oscillations
of the current $I_{c}$ {\bf \ }as a function of the F layer thickness $%
2d_{F} $ or temperature, observed in experiments.
\end{abstract}

\section{Introduction}

\bigskip

New states have been and observed in Josephson junctions in the last years.
These states are characterized by a negative Josephson energy $%
E_{J}=(I_{c}\hbar /2e)(1-\cos \phi ),$ that is, under certain conditions the
Josephson critical current $I_{c}$ changes its sign becoming negative. This
means that the ground state corresponds to the phase difference $\phi $
equal to $\pi $ ($\pi $-state)$,$ but not to zero as it takes place in
ordinary Josephson junctions. Such states have been observed in Josephson
junctions of different types:1) in multiterminal SNS Josephson junctions 
\cite{Wees}, 2) in junctions consisting of two $d$-wave superconductors (see
references in the review \cite{Kirtley}), 3) in SFS junctions \cite
{Ryazanov,Kontos,Blum} (see also the review \cite{GolubovRev}), where S,N
and F stand for a superconductor, a normal metal and a ferromagnet,
respectively. The $\pi $-state in SNS junctions is created by passing a
dissipative current through the N layer. This current leads to a
nonequilibrium distribution function of quasiparticles in the N wire with
respect to the equilibrium distribution function in the superconductors. In
the $d$-wave superconductors (high $T_{c}$-superconductors) the sign of the
order parameter $\Delta $ depends on the direction in space with respect to
crystallographic axes. Therefore if two S/N (or S/I, where I is an isolator)
interfaces have different, properly chosen orientations, then the critical
current $I_{c},$ which is proportional to the product $\Delta _{1}\Delta
_{2},$ may be negative. This occurs provided that the order parameter in one
superconductor ($\Delta _{1}$ or $\Delta _{2}$) is negative. In SFS
junctions the critical current $I_{c}$ may be negative because the
condensate (Gor'kov's) Green's function, which determines the current $I_{c}$
(or to be more exact, the critical current density $j_{c}$)$,$ oscillates in
space changing sign. The sign reversal of $j_{c}$ in SFS junctions has been
predicted a long time ago by Bulaevskii, Kuzii and Sobyanin \cite{Bul77},
who considered electron tunneling between two superconductors via a magnetic
impurity. Later this effect was studied in SFS junctions by Buzdin,
Bulaevskii and Panyukov \cite{Buzdin82} (references to other theoretical
papers on this subject is given in the review \cite{GolubovRev}). It was
shown that the current density $j_{c}$ decays with increasing the thickness
of the F layer $2d_{F}$ and changes sign (damped oscillations of $j_{c}$).
Such a behaviour of the critical current has been observed experimentally in
Refs. \cite{Ryazanov,Kontos,Blum,Strunk}; the current $I_{c}$ decays with
increasing the thickness $2d_{F}$ or temperature $T$ in a nonmonotonic way
changing sign. In the recent paper \cite{Strunk} a spontaniously circulating
current in a superconducting ring with a SFS $\pi $-junction has been
observed.

In this paper we show that the damped $I_{c}$ oscillations are not
necessarily related to such a dependence of the local critical current
density $j_{c}$ as it was anticipated previously. The total Josephson
current $I_{J}$, which is measured in experiments, is an integral from the
local current density $j_{J}=$ $j_{c}\sin \phi $ over the whole area of a
SFS junction (we choose a simple, sinusoidal form of \ the dependence $%
j_{J}(\phi ),$ but the conclusions we make are valid qualitatively in a
general case). It is important to have in mind that the phase difference $%
\phi $ varies in space in the presence of a magnetic field, and in
multidomain SFS junctions a spatial dependence of $\phi $ arises even in the
absence of an external magnetic field $H_{ext}$. It was already mentioned in
Ref.\cite{Ryazanov,Kontos} that the maximum value of $I_{c}$ corresponds to
zero external magnetic field $H_{ext}.$ One can assume that this may be
related to a multidomain structure of the F film. Otherwise the maximum
value of $I_{c}$ would be shifted by a certain value of $H_{ext}$ for which
the induction $H_{ext}+4\pi M$ is zero, where $M$ is the magnetization in a
one-domain F film. We consider a simple model of a multidomain structure of
the F film and show that even if the local current density $j_{c}$ is always
positive, the critical current $I_{c}$ changes sign when the in-plane
magnetic flux in a domain $\Phi _{a}=4\pi M_{0}(2d_{F}a)$ is an integer of
the flux quantum $\Phi _{0},$ where $a$ is the domain width. The domain
width $a$ depends both on the thickness $d_{F}$ and temperature $T$ if the
screening of stray magnetic fields by the Meissner currents is taken into
account \cite{Taras,Bulaev,Sonin}. Therefore the magnetic flux is changed
with increasing $d_{F}$ and temperature leading to damped oscillations of
the critical current $I_{c}.$ It will be shown that the dependence $%
I_{c}(\Phi _{a}/\Phi _{0})$ may be described by a Fraunhofer-like pattern

\begin{equation}
I_{c}(\Phi _{a}/\Phi _{0})=I_{c0}\sin (\pi \Phi _{a}/\Phi _{0})/(\pi \Phi
_{a}/\Phi _{0})  \label{Fraun}
\end{equation}
where $I_{c0}=j_{c}L_{x}$ is the critical current for an uniform
(one-domain) SFS junction per unit length in the $y$-direction (we assume
that all quantities depend only on $x$ and $z;$ see Fig.1). This dependence
describes the critical current for the case of domains with an alternating
magnetization which takes constant values $\pm M_{0}$ in domains.

\section{\protect\bigskip Model and basic equations}

Consider a SFS Josephson junction in which the F layer consists of stripe
domains parallel to the $y$-axis (see Fig.1).

\begin{figure}[tbp]
\epsfysize= 8cm \vspace{0.2cm}
\centerline{\epsfbox{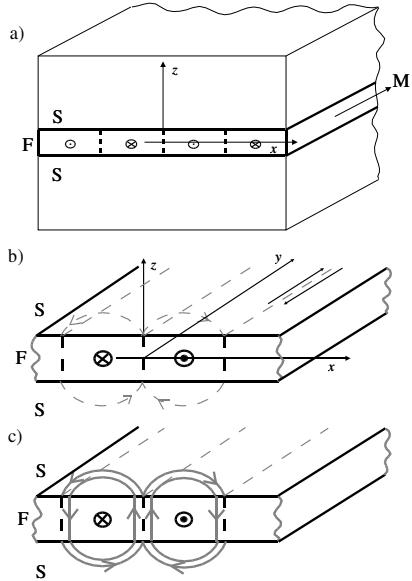}} \vspace{0.2cm}
\large{\caption{Schematic picture of the considered SFS junction with magnetic domains (1a).  Fig.1b
shows the supercurrents (solid lines with arrows) in the upper
superconductor screening the normal component of a magnetic field near domain walls. These currents in the lower superconductor (not shown) have
the same direction. The domain boundaries are represented by the dashed
lines. The dashed lines with arrows illustrate the stray magnetic field
related to rotation of the magnetization ${\bf M}$ in the domain walls. The cross and  dot in the circles show the directions of the magnetization vector in neighboring domains. 
In Fig.1c the solid lines with arrows show the screening supercurrents in
presence of the Josephson coupling.}}

\end{figure}

To be specific, we consider the Bloch domains, that is, the magnetization
vector ${\bf M}$ lies in the $y-z$ plane changing its direction in domain
walls of the width $w$. In the case of a narrow domain width the
magnetization vector $M$\ is parallel to the $y$-axis, constant in each
domain and rotates in the domain wall. The magnitude $|M|$\ is assumed to be
constant everywhere in the F film. The orientation of the
magnetization $M$\ in the F film (in-plane or out-of-plane) depends on many
factors such as the Curie temperature, the constant of magnetic anisotropy,
the film thickness etc (see for example, \cite{Levan,Pokr} and references
therein). In ultrathin F films (a few atomic layers) a transition between
the in-plane to out-of-plane $M$\ orientation may occur \cite{Pappas}. We
consider much more thicker F films, used in experiments \cite
{Ryazanov,Kontos,Blum,Strunk}, where the magnetization orientation is
determined by the direction of easy magnetization. Therefore it is assumed
that the axis of easy magnetization is parallel to the plane of the F film.
Even in the absence of the Josephson coupling between the superconductors S,
the Meissner currents along the $y$-direction are induced by the stray
magnetic field ${\bf H=}$ ${\bf (}H_{x},0,H_{z}{\bf )}$ (dashed lines with
arrows in Fig. 1a)${\bf .}$ The magnetic field components $H_{x,z}$ can be
easily found from equations

\begin{equation}
\partial ^{2}H_{x,z}(k_{n},z)/\partial z^{2}-\kappa
_{n}^{2}H_{x,z}(k_{n},z)=0  \label{Hxz}
\end{equation}
where $H_{x,z}(k_{n},z)=\int_{-a}^{a}(dx/2a)H_{x,z}(x,z)\exp (ik_{n}x)$ is
the Fourier component of $H_{x,z}(x,z)$, $k_{n}=\pi n/a,$ $n=0,\pm
1,...;\kappa _{n}^{2}=k_{n}^{2}+\kappa _{L}^{2}$, $\kappa _{L}^{-1}=\lambda
_{L}$ is the London penetration depth (in the ferromagnet the penetration
depth $\lambda _{L,F}$ may be taken infinite because the amplitude of the
condensate in the F layer is small and therefore the screening is weak).
This equation is supplemented by the boundary conditions \cite{LL}

\begin{equation}
\lbrack H_{x}]=0,\text{ }[H_{z}]=4\pi M_{z}(k_{n},z)  \label{BC}
\end{equation}
where the square brackets mean a difference: $%
[H_{x}]=H_{x}(k_{n},d_{F}+0)-H_{x}(k_{n},d_{F}-0).$ The component $M_{z}$\
is not zero in the domain walls. This implies that the in-plane component of 
$H$ is continuous across the S/F boundary and the normal component of $H,$
which exists near the domain walls, experiences a jump. One can easily solve
Eqs.(\ref{Hxz}) and {find the fields }$H_{x,z}$ which are connected with the
vector potential ${\bf A:}$ $H_{x}=-\partial A_{y}/\partial z,$ $%
H_{z}=\partial A_{y}/\partial x$. The expression for\ $A_{y}$ in the
superconductor is

\begin{equation}
A_{y}(k_{n},z)=-4\pi M_{z}(k_{n})(f_{n}ik_{n})^{-1}\exp (-\kappa
_{n}(z-d_{F}));\text{ }z>d_{F}  \label{Ay}
\end{equation}
where $f_{n}=1+\kappa _{n}/(k_{n}\tanh \theta _{n}),\theta _{n}=k_{n}d_{F}.$
The vector potential $A_{y}$ in the lower superconductor ($z<d_{F}$) has the
same sign. This means that the stray fields $H_{x,z}$ lead to the screening
currents

\begin{equation}
j_{y}(k_{n},z)=-(c/4\pi )\kappa _{L}^{2}A_{y}(k_{n},z)  \label{jy}
\end{equation}
which have the same direction in the upper and lower superconductors, but
the opposite directions in neighboring domains (see Fig.1a). Because the
Meissner currents flow in the same direction in both superconducting
electrodes, they do not lead to a phase difference between the
superconductors and do not inluence the Josephson current essentially. These
currents may locally reduce the amplitude of the order parameter and
therefore decrease the critical current $I_{c}$ if the magnetization $M_{z}$
is strong enough. We will not discuss this simple effect.

If the Josephson coupling between the superconductors is negligible (this
case was considered in Ref.\cite{Taras}), the magnetic field has only the
components $H_{x,z}$ which are determined by the vector potential $%
A_{y}(k_{n},z)${. The component }$H_{y}$ is zero. However the components $%
A_{x,z}(k_{n},z)$ of the vector potential are not zero. For example in
one-domain case

\begin{eqnarray}
A_{x}(z) &=&4\pi M_{y}z\text{ in F}  \label{Ax} \\
A_{x}(z) &=&\pm 4\pi M_{y}d_{F}\text{ in S}
\end{eqnarray}

In the absence of the Josephson current the components $A_{x,z}$ in a
multidomain SFS structure are found from the equations: ${\bf H=\nabla
\times A}=0$ and ${\bf \nabla \cdot A}=0.$ However in this case the presence
of components $A_{x,z}$ does not lead to the superconducting currents and
therefore to the appearance of an additional magnetic field $H.$ The term
proportional to $A_{x}$ in the expression for the supercurrent is
compensated by the term proportional to gradient of the phase (see Eq. ({\ref
{Max1}})). {If the Josephson current }$j_{J}$ is not zero, additional
screening currents $j_{x,z}$ and therefore the component $H_{y}$ arise in
the system (see Fig.1b). The component $A_{x}$ affects the phase difference
and therefore the critical current $I_{c}$ in the junction under
consideration.

In order to calculate the critical current $I_{c},$ one needs to derive an
equation governing the phase difference $\phi $ in a multidomain SFS
Josephson junction. For simplicity we assume that the thickness of the F
layer is small: $2d_{F}<<a,w$, where $2a$ is the period of the domain
structure and $w$ is the width of the domain wall (one can analyze a more
general case, but the calculations become more combersome). We need an
equation for the component $H_{y}$ which is related to the local Josephson
current density

\begin{equation}
(\nabla \times H)_{z}=\partial H_{y}/\partial x|_{z=0}=(4\pi /c)j_{c}\sin
\phi  \label{rotHz}
\end{equation}

We assumed the simplest form of the relationship between the Josephson
current density $j_{J}$ and the phase difference $\phi ,$ but this
assumption is not essential. We also dropped a contribution to the
gauge-invariant ''phase difference'' which stems from the vector potential
and has the form $\int_{-d_{F}}^{d_{F}}A_{z}dz$. One can easily show that
this part is smaller than $\phi $ by the parameter $(a/d_{F})$ (we choose a
gauge in which ${\bf \nabla \cdot A}=$ $0$). Let's write down the $x$%
-component of one of the Maxwell equations for the current density in the
superconductor at $z=\pm d_{F}$

\begin{equation}
-\partial H_{y}/\partial z=(4\pi /c)j_{x}=\kappa _{L}^{2}(-A_{x}+(\Phi
_{0}/2\pi )\partial \chi /\partial x)  \label{Max1}
\end{equation}
where $\kappa _{L}^{2}$ is defined in Eqs.{(\ref{Hxz}) and }$\chi $ is the
phase of the order parameter. Subtract Eqs.{(\ref{Max1}) taken at } $%
z=+d_{F} $ and $z=-d_{F}$ from one another (a similar method was used by one
of the authors in Ref.\cite{VolkovPL} in the study of collective modes in
layered superconductors), we get

\begin{equation}
-[\partial H_{y}/\partial z]=\kappa _{L}^{2}(-[A_{x}]+(\Phi _{0}/2\pi
)\partial \phi /\partial x)  \label{Max2}
\end{equation}
where the square brackets means, as before, a jump across the F layer and $%
\phi =\chi (d_{F})-\chi (-d_{F})$ is the phase difference. The jump $[A_{x}]$
is found from the equation

\begin{equation}
4\pi M_{y}=\partial A_{x}/\partial z-\partial A_{z}/\partial x  \label{My}
\end{equation}
and is equal to: $[A_{x}(k_{n})]=4\pi M_{y}(k_{n})d_{F}$ in accordance with
Eq.({\ref{Ax})(see Footnote \cite{Foot})}. The contribution of the second
term is smaller by the parameter $(d_{F}/a).$ The component of the magnetic
field $H_{y}$ can be found from an equation similar to Eqs.{(\ref{Hxz}).
With account for the boundary condition (\ref{Max2}) this equation acquires
the form}

\begin{equation}
\partial ^{2}H_{y}(k_{n},z)/\partial z^{2}-\kappa
_{n}^{2}H_{y}(k_{n},z)=\delta (z)\kappa _{L}^{2}([A_{x,k}]+(\Phi _{0}/2\pi
)ik_{n}\phi _{k})  \label{Hy}
\end{equation}

Solving this equation for $H_{y}$ and substituting the solution into Eq.{(%
\ref{rotHz}), we obtain an equation for Fourier components of the phase
difference }$\phi _{k}=\phi (k_{n},z)$

\begin{equation}
\kappa _{n}^{2}\phi _{k}+\kappa _{J}^{2}(\kappa _{n}/\kappa _{L})(\sin \phi
(x))_{k}=ik_{n}4\pi M_{y}(k_{n})(2d_{F})(2\pi /\Phi _{0})  \label{PhiK}
\end{equation}
where $\kappa _{J}=\sqrt{16\pi ^{2}j_{c}/c\kappa _{L}\Phi _{0}}$ is the
inverse Josephson length. In the coordinate representation Eq.{(\ref{PhiK})
has the form } 
\begin{equation}
-\partial ^{2}\phi /\partial x^{2}+\kappa _{J}^{2}\int dx_{1}K(x-x_{1})\sin
\phi (x_{1})=-4\pi d_{F}(2\pi /\Phi _{0})\partial M_{y}(x)/\partial x
\label{PhiS}
\end{equation}
where the kernel $K(x-x_{1})$ is defined as follows

\begin{equation}
K(x-x_{1})=(1/2a)\sum_{n}(\kappa _{n}/\kappa _{L})\exp (-ik_{n}(x-x_{1}))
\label{K}
\end{equation}
Eq.{(\ref{PhiS}) describes the dc Josephson effect in a simple model of
multidomain SFS junctions with a thin F layer.}

\section{Two types of domain structures}

\bigskip If the London penetration depth is small compared to the domain
size $a$ and the width of the domain wall ( $\lambda _{L}<<a,w$ ), then Eq.({%
\ref{PhiS}}) is simplified$.$ This condition means that the characteristic $%
k_{n}$ values are much smaller than $\kappa _{L}$. In this case $%
K(x-x_{1})\approx \delta (x-x_{1})$ and Eq.{(\ref{PhiS}) acquires the form}

\begin{equation}
-\partial ^{2}\phi /\partial x^{2}+\kappa _{J}^{2}\sin \phi (x)=-4\pi
(2d_{F})(2\pi /\Phi _{0})\partial M_{y}(x)/\partial x  \label{PhiLoc}
\end{equation}
This equation differs from the standard Josephson equation only by the term
on the right-hand side. One can study various properties of the SFS
junctions described by Eq.({\ref{PhiLoc}}) or by Eq.({\ref{PhiS}}), but in
this paper we analyze only the critical current and its dependens on
different parameters (external magnetic field, the thichkness $d_{F}$ etc).
First we consider the case of a periodic $M_{y}(x)$ dependence. If the
period $2a$ of this dependence is much less than the long Josephson length $%
\kappa _{J}^{-1},$ then a solution for Eq.{(\ref{PhiS}) (the relation
between }$\lambda _{L}$ and $a,w$ may be arbitrary{) is}

\begin{equation}
\phi =\phi _{M}+\phi _{0},\text{ \ }\phi _{M}=(2\pi /\Phi _{0})4\pi
(2d_{F})\int^{x}dx_{1}M_{y}(x_{1})  \label{Phi}
\end{equation}
where $\phi _{M}(x)$ is a function fast varying in space, $\phi _{0}$ is a
constant (or a function smoothly varying over the period $a$). The total
Josephson current (per unit length in $y$-direction) is

\begin{equation}
I_{J}=j_{c}\int_{0}^{L_{x}}dx_{1}\sin (\phi _{0}+\phi _{M}(x_{1}))
\label{IJ}
\end{equation}
where $\phi _{M}$ is given by Eq.({\ref{Phi}}). Note that the weak Josephson
coupling does not affect the domain structure, and therefore this structure
can studied in the absence of the Josephson effect. As we noted, the domain
structure was analyzed theoretically in Ref.\cite{Taras} for arbitrary $%
d_{F} $ and in Refs.\cite{Bulaev,Sonin} for thick F layers ($d_{F}>>a$). It
was shown in Ref.\cite{Taras} that the period $a$ depends on $d_{F}$ in a
nonmonotonic way and the width of the domain walls $w$ may be much less than
or comparable with the domain width $a$. One has a step-like structure $%
M_{y}(x)$ in the first case and an oscillatory structure in the second case.
Consider two limiting cases.

\begin{figure}[tbp]
\epsfysize= 10cm \vspace{0.2cm}
\centerline{\epsfbox{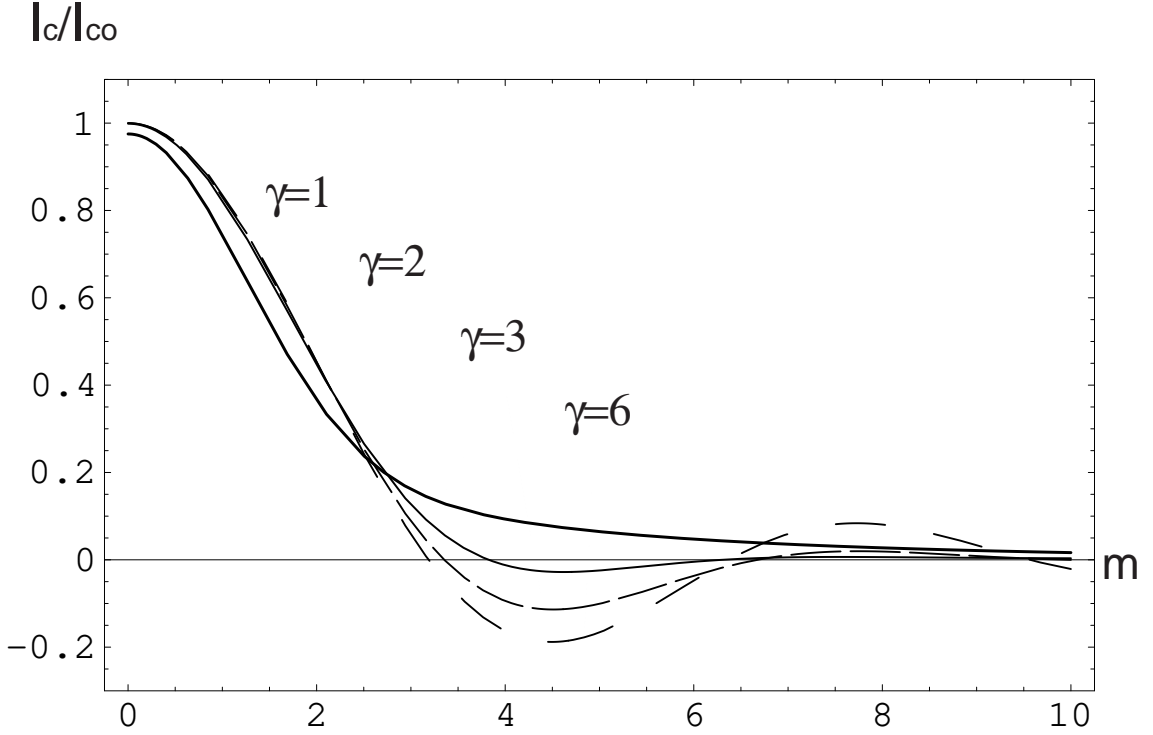}} \vspace{0.2cm}

\large{\caption{Normalized critical Josephson current as a function of the
normalized magnetic flux in a domain $m=\protect\pi \Phi _{a}(a)/\Phi _{0}$
for different parameters $\protect\gamma =\overline{a}/\protect\delta $,
where $\overline{a}$ is the averaged domain size and $\protect\delta/\sqrt{2} $ is
the dispersion of the domain size fluctuations.}}
\end{figure}

a) Step-like domain structure. The magnetization vector equals:$%
M=(0,M_{y}(x),0),$\ where $M_{y}(x)$\ $=$\ $+M_{0}$\ for $0<x<a,$\ and $%
M_{y}(x)$\ $=-M_{0}$\ for $-a<x<0.$ Outside this interval the dependence $%
M_{y}(x)$ is periodically repeated. For this structure the Josephson current 
$I_{J}$ is described by the formula

\begin{equation}
I_{J}=I_{c0}\sin \phi _{0}\frac{\sin (\pi \Phi _{a}/\Phi _{0})}{\pi \Phi
_{a}/\Phi _{0}}  \label{IJstep}
\end{equation}
where $I_{c0}=j_{c}L_{x}$, $\Phi _{a}=4\pi M_{0}(2d_{F}a)$ is the in-plane
magnetic flux in one domain$.$ Therefore the critical current $I_{c}$, which
is given by Eq.({\ref{Fraun}}) oscillates and decays with increasing $\Phi
_{a}.$ This implies that the amplitude of the critical current oscillations
decreases with increasing $d_{F}$ or temperature $T$ because the period of
the domain structure $2a$ depends on $\lambda _{L}(T)$ (see the theoretical
papers \cite{Taras,Bulaev,Sonin} and the experimental paper \cite{Geim},
where it was shown that the domain structure is changed with changing $T$).

b) Oscillatory domain structure: $M_{y}(x)$ = $M_{0}\sin (k_{0}x),$ $%
k_{0}=\pi /a.$

In this case the critical current is equal to

\begin{equation}
I_{c}=I_{c0}J_{0}(\phi _{M}(a))  \label{IJosc}
\end{equation}
where $J_{0}$ is the Bessel function of the zeroth order. In both cases the
behaviour of the critical current $I_{c}$ as a function of $\Phi _{a}$ is
qualitatively the same: the current $I_{c}$ decreases with increasing $\phi
_{M}(a)$ and changes sign.

In our model of a periodic domain structure the action of an in-plane
external magnetic field $H_{ext}$ on $I_{c}$ can be easily analyzed. In the
presence of the field $H_{ext}$ the phase difference equals $\phi (x)=\phi
_{0}+\phi _{H}(x)+\phi _{M}(x)$, where $\phi _{M}(x)$ is given by Eq.({\ref
{Phi}}), $\phi _{0}$ is a constant and $\phi _{H}(x)=2\lambda
_{L}xH_{ext}(2\pi /\Phi _{0}).$ If the domain size $a$ is much less than the
length of the junction $L_{x}$, the averaging over the period of the
structure can be done as before at a fixed coordinate $x$, and we arrive at
Eq.({\ref{IJstep}}) in which one has to replace $\phi _{0}\Longrightarrow
\phi _{0}+\phi _{H}(x).$ The final averaging over $L_{x}$ yields, for
example, in the model of a step-like domain structure for $I_{c}$

\begin{equation}
I_{c}(H_{ext})=I_{c0}\frac{\sin \phi _{H}(L_{x})}{\phi _{H}(L_{x})}\frac{%
\sin (\pi \Phi _{a}/\Phi _{0})}{\pi \Phi _{a}/\Phi _{0}}  \label{IcH}
\end{equation}
Thus the dependence $I_{c}(H_{ext})$ is given by the usual Fraunhofer curve
with an effective critical current $I_{c0}\sin (\pi \Phi _{a}/\Phi
_{0})/(\pi \Phi _{a}/\Phi _{0})$ the sign and value of which depends on $%
d_{F},a$ etc.

Although in the theoretical papers \cite{Taras,Bulaev,Sonin} only a regular
domain structure is considered, in real samples the domain structure is not
strictly periodic. It may be almost regular (see for example, \cite
{Beasley,Ved}), or very irregular (\cite{Speck,Lange}) with in-plane or
out-of plane magnetizations. We study the effect of a possible irregularity
of the domain structure on the basis of a simple model. We assume simply
that the domain size $a$ fluctuates around a mean value $\overline{a}$ and
fluctuations of $a$ are described by the Gaussian distribution. Then the
dependence of $I_{c}(\delta )$ on the dispersion of the fluctuations is
given by the integral (in the absence of $H_{ext}$)

\begin{equation}
I_{c}(\delta )=I_{c0}c_{1}\int_{0}^{\infty }da\frac{\sin (\pi \Phi
_{a}(a)/\Phi _{0})}{\pi \Phi _{a}(a)/\Phi _{0}}\exp (-(a-\overline{a}%
)^{2}/\delta ^{2})  \label{IcFluc}
\end{equation}

where $c_{1}=(\int_{0}^{\infty }da\exp (-(a-\overline{a})^{2}/\delta
^{2}))^{-1}$ is a normalization constant. In Fig.2 we plot the dependence $%
I_{c}(\pi \Phi _{a}(\overline{a})/\Phi _{0})$ for different parameter $%
\gamma =\overline{a}/\delta $. One can see that for large $\gamma $ this
dependence coincides with a Fraunhofer pattern, but with decreasing $\gamma $
the amplitude of oscillations of $I_{c}$ decreases and finally the function $%
I_{c}(\Phi _{0})$ does not change sign (no $\pi -$states).

\section{Conclusion}

\bigskip

In conclusion, using a simple model of a multidomain SFS Josephson junction,
we have calculated the critical current $I_{c}$. It turns out that the
current $I_{c}$ changes signs when the in-plane magnetic flux $\Phi
_{a}=4\pi Ma(2d_{F})$ in each domain equals $n\Phi _{0}.$ The magnetic flux $%
\Phi _{a}$ is caused by the magnetization in the ferromagnetic domains and
therefore exists even in the absence of an external magnetic field. The
oscillations of $I_{c}$ observed experimentally by varying thickness $2d_{F}$
or temperature $T$ may be related not to the sign reversl of the local
critical current density $j_{c}$, but to a simple mechanism - the
Fraunhofer-like oscillations of $I_{c}$ caused by the internal magnetization 
$M$ in domains. Almost nothing is known about the domain structure in real
SFS junctions. For estimations we take $4\pi M_{0}\approx 1kOe,$ $a\approx
1mkm,$ $2d_{F}\approx 100A$. For these values we obtain $\Phi _{a}\approx
10^{-7}Oe\cdot cm^{2}$. This means that the critical current $I_{c}$ changes
sign for the thickness $2d_{F}$ about $100A.$ This value of thickness is
close to that used in experiments, although, strictly speaking, the values
of the magnetization $M$ and of the domain structure period $a$ are not
known. In order to make more convincing conclusions about what is the
mechanism of the sign reversal effect (whether it is caused by the sign
reversal of the critical current density $j_{c}$ or by the spatial phase
modulation in a multidomain SFS structure), further theoretical and,
especially, experimental studies are needed. In particular, it would be
interesting to study the influence of the domain structure on Shapiro steps
in SFS junctions measured in a recent paper \cite{Sellier}

We would like to thank SFB 491 for a financial support.

\bigskip

\bigskip


\begin{references}
\bibitem{Wees}  J. J. A. Baselmans, A. F. Morpurgo, B. J. van Wees, and T.
M. Klapwijk, Nature (London) {\bf 397}, 43 (1999).

\bibitem{Kirtley}  C. C. Tsuei, and J. R. Kirtley, Rev. Mod. Phys. {\bf 72},
969 (2000).

\bibitem{Ryazanov}  V. V. Ryazanov, V. A. Oboznov, A. Yu. Rusanov, A. V.
Veretennikov, A. A. Golubov, and J. Aarts, Phys. Rev. Lett. {\bf \ 86}, 2427
(2001).

\bibitem{Kontos}  T. Kontos, M. Aprili, J. Lesueur, F. Gen\^{e}t, B.
Stephanidis, and R. Boursier,

Phys. Rev. Lett {\bf 89}, 137007 (2002).

\bibitem{Blum}  Y. Blum, A. Tsukernik, M. Karpovski, and A. Palevski, Phys.
Rev. Lett. {\bf 89}, 187004 (2002).

\bibitem{Strunk}  A. Bauer, J. Bentner, M. Aprili, M. L. Della Rocca, M.
Reinwald, W. Wegscheider, and C. Strunk, Phys. Rev. Lett. {\bf 92}, 217001
(2004).

\bibitem{GolubovRev}  A.A. Golubov, M.Yu. Kupriyanov, and E. Il'ichev,
Rev.Mod. Phys. {\bf 76, }411 (2004).

\bibitem{Bul77}  L.N.Bulaevskii, V.V.Kuzii, and A.A.Sobyanin, Sov. Phys.
JETP Lett. {\bf 25}, 299 (1977).

\bibitem{Buzdin82}  A.I.Buzdin, L.N.Bulaevskii, and S.V.Panyukov, Sov. Phys.
JETP Lett. {\bf 35}, 178 (1982).

\bibitem{Taras}  A.Stankiewicz, S.\ J.\ Robinson, G. A. Gehring, and V. V.
Tarasenko, J.Phys. Condens. Matter {\bf 9}, 1019 (1997).

\bibitem{Bulaev}  L.N. Bulaevskii and E.M. Chudnovsky, Phys. Rev. B {\bf 63}%
, 012502 (2001).

\bibitem{Sonin}  E.B. Sonin, Phys. Rev. B {\bf 66},136501 (2002).

\bibitem{Levan}  A. P. Levanyuk and N. Garcia, Phys. Rev. Lett. {\bf 70},
1184 (1993).

\bibitem{Pokr}  A. B. Kashuba and V. L. Pokrovsky, Phys. Rev.{\bf \ B 48, }%
10335 (1993){\bf .}

\bibitem{Pappas}  D. P. Pappas, K.P. K\"{a}mper, and H. Hopster, Phys. Rev.
Lett. {\bf 64}, 3179 (1990)

\bibitem{LL}  L.D.Landau and E.M.Lifshitz, {\it Electrodynamics of
Continuous Media,} Butterworth-Heinemann, Oxford, 1982, p. 107.

\bibitem{Geim}  S. V. Dubonos, A. K. Geim, K. S. Novoselov, and I. V.
Grigorieva, Phys. Rev.{\bf \ B 65}, 220513 (2002)

\bibitem{VolkovPL}  A.F. Volkov, Phys. Lett., {\bf A 138}, 213 (1989).

\bibitem{Foot}  Note that an induced magnetization arises also in the
superconductors. As shown in Refs. (F.S. Bergeret, A. F. Volkov, and K. B.
Efetov, B {\bf 69}, 174504 (2004);{\bf \ }Europhys.Lett{\bf . 66,} 111{\bf \ 
}(2004){\bf \ }), the magnetization penetrates from the ferromagnet into the
superconductor (with the opposite sign) over the correlation length $\xi
_{S} $\ if the proximity effect is essential. In the case of an itinerant
ferromagnet (the magnetization is mainly due to free electrons) the total
magnetic moment in the ferromagnet may be completely compensated by the
magnetic moment induced in the superconductor. We do not take into account
this inverse proximity effect assuming that $M_{F}$ is produced mainly by
localized magnetic moments. Account for the inverse proximity effect leads
to replacement of $M_{y}$ by an effective magnetization $M_{y,eff}.$

\bibitem{Sellier}  H. Sellier, C. Baraduc, F. Lefloch, and R. Calemczuk,
Phys. Rev. Lett. {\bf 92}, 257005 (2004).

\bibitem{Beasley}  M. Feigenson, L. Klein, J. W. Reiner, and M. R. Beasley,
Phys. Rev. {\bf B 67}, 134436 (2003).

\bibitem{Ved}  E. Y. Vedmedenko, A. Kubetzka, K. von Bergmann, O. Pietzsch,
M. Bode, J. Kirschner, H. P. Oepen, and R. Wiesendanger, Phys. Rev. Lett. 
{\bf 92}, 077207 (2004).

\bibitem{Speck}  M. Speckmann, H. P. Oepen, and H. Ibach, Phys. Rev. Lett. 
{\bf 75}, 2035 (1995).

\bibitem{Lange}  M. Lange, M. J. Van Bael, and V. V. Moshchalkov, Phys. Rev. 
{\bf B 68}, 174522 (2003)
\end{references}
\end{document}